\documentclass[10pt]{article}

\usepackage{amsmath}
\usepackage{graphicx}
\usepackage{indentfirst}
\usepackage{subfigure}
\usepackage{hyperref}

\hypersetup{
  colorlinks	= true,	
  urlcolor		= blue,	
  linkcolor		= blue,	
  citecolor		= red	
}

\begin{document}

\title{\bf{Spin Interaction under the Collision of Two Kerr-(anti-)de Sitter Black Holes}}

\date{}
\maketitle

\begin{center}
\author{Bogeun Gwak}$^a$\footnote{rasenis@sejong.ac.kr} and \author{Daeho Ro}$^b$\footnote{daeho.ro@apctp.org }\\

\vskip 0.25in
$^{a}$\it{Department of Physics and Astronomy, Sejong University, Seoul 05006, Republic of Korea}\\
$^{b}$\it{Asia Pacific Center for Theoretical Physics, POSTECH, Pohang, Gyeongbuk 37673, Republic of Korea}\\
\end{center}
\vskip 0.6in

{\abstract
{
We have investigated spin interaction under the collision of Kerr-(anti-)de Sitter black holes. The potential of a spin interaction is dependent on the relative rotating directions of the black holes, and this potential can be released as gravitational radiation under the collision. The radiation depends on the cosmological constant and corresponds to the potential of the spin interaction at a limit where one of the black holes is assumed to have small mass and angular momentum. Then, we have shown, approximately, the overall behaviors of the upper bounds on the radiation using thermodynamics. From these bounds, the spin interaction can consistently contribute to the radiation. In addition, the radiation depends on the stability of the black hole synthesized from the collision.
}}

\thispagestyle{empty}
\newpage
\setcounter{page}{1}
\section{Introduction}

As recently observed in the Laser Interferometer Gravitational-Wave Observatory (LIGO), gravitational waves are generated by a binary black hole merger\cite{Abbott:2016blz,Abbott:2016nmj}. In particular, for the GW150914 at the first detection of the gravitational wave\cite{Abbott:2016blz}, the black holes are massive, and their masses are more than ten times that of the solar mass. The energy of the gravitational radiation is approximately three times that of the solar mass so that the collision of black holes is an important source of gravitational radiation that is detectable in our universe. Another important observation is about the value of the cosmological constant. Our universe is now in accelerated expansion, so the cosmological constant should be positive. According to several astrophysical observations\cite{Perlmutter:1996ds,Caldwell:1997ii,Garnavich:1998th}, the value of the cosmological constant is definitely small. In the theory of Einstein gravity with a positive cosmological constant, the solution for the black hole is the de Sitter (dS) black hole. Therefore, the dS black hole is an appropriate to description for phenomena occurred in our universe. On the contrary, the negative sign of the curvature is provided by anti-de Sitter (AdS) spacetime. AdS spacetime itself is distant from our universe, but it plays an important role in gauge/gravity duality called anti-de Sitter/conformal field theory (AdS/CFT) duality\cite{Witten:1998zw,Maldacena:1997re}. The theory of gravity in $D$-dimensional AdS spacetime corresponds with the $(D-1)$-dimensional CFT on the AdS boundary due to the AdS/CFT duality. Dual CFT can be described in terms of a finite temperature from the AdS black hole\cite{Gubser:1998bc,Witten:1998qj,Aharony:1999ti} because thermal properties, such as temperature, can appear at the boundary of AdS spacetime. In the progress of this duality, an interesting correspondence for AdS spacetime can be found in condensed matter theory (CMT). This AdS/CMT correspondence as well as the holographic superconductor given from the correspondence can be treated using charged AdS black holes\cite{Chaturvedi:2013ova,Haehl:2013hoa,Chang:2014jna}.

The energy of a black hole can be extracted from a particle through the Penrose process\cite{Bardeen:1970zz,Penrose:1971uk}. Under this process, the black hole only loses reducible energy, such as rotational and electric energies, but another type of energy, called irreducible mass, always increases even with the Penrose process\cite{Christodoulou:1970wf,Christodoulou:1972kt}. This irreducible energy is distributed on the surface of the horizon\cite{Smarr:1972kt} and is proportional to the square root of the area of the horizon. As the square of the irreducible mass, the area of the horizon is a kind of properties that increase by adding an external particle or field. From this point, the area is similar to the entropy of the thermal system, so that the Bekenstein-Hawking entropy of the black hole can be defined as the area of the horizon\cite{Bekenstein:1973ur,Bekenstein:1974ax}. In addition, the black hole can radiate energy through the quantum effect from which the Hawking temperature of the black hole can be provided with a temperature similar to that of the black body\cite{Hawking:1974sw,Hawking:1976de}. Hence, the black hole can be treated as a thermal system with temperature and entropy, and the laws of thermodynamics can be also defined in terms of its thermal properties.

The gravitational radiation released from the collision of black holes has an upper limit obtained from the laws of thermodynamics\cite{Hawking:1971tu}, but this upper limit is large in the collision of identical black holes. More precise results can be provided when we assume a high-energy collision between the black holes\cite{Eardley:2002re,Sperhake:2008ga,Coelho:2012sya}. The collision of black holes and the waveform of the gravitational radiation can be shown by numerical relativity in the theory of Einstein gravity\cite{Smarr:1976qy,Smarr:1977fy,Smarr:1977uf,Witek:2010xi}. At present, the waveform generated from the collision of black holes can be investigated under various initial conditions\cite{Bantilan:2014sra,Bednarek:2015dga,Hirotani:2015fxp,Sperhake:2015siy,Barkett:2015wia,Hinderer:2016eia}. In addition, the gravitational radiation depends on the relative rotations of the black holes due to their spin interaction, which has an effect even if the rotating planes of the black holes are parallel. The potential of the interaction is negative for an anti-parallel arrangement of rotating planes but positive for a parallel one. Therefore, black holes become attractive for anti-parallel arrangements and repulsive for parallel ones\cite{Schiff:1960gi,Wilkins:1970wap,Mashhoon:1971nm,Wald:1972sz}. The spin interaction can appear to be the spinning particle moving in spacetime\cite{Plyatsko:2015bia,d'Ambrosi:2015xci}. Since the spin interaction increases along with the mass and angular momentum of the black hole system, this interaction can be large enough to affect the stability of a binary black hole\cite{Majar:2012fa,Zilhao:2013nda}.

The stability of black holes is studied using various methods. For the asymptotically flat case, the Kerr black hole is stable in perturbation\cite{Teukolsky:1972my,Press:1973zz}, but the energy of this black hole can be extracted by an external field. Hence, if the Kerr black hole is in a cavity, the external field can continuously take the energy of the black hole, so the black hole can be unstable under this process. This phenomenon is called superradiance\cite{Press:1972zz}. Physically, the Kerr black hole in a cavity is similar to a Kerr-AdS black hole because the AdS boundary reflects the external field. Therefore, the AdS black hole has the instability of superradiance\cite{Cardoso:2004hs,Cardoso:2006wa,Cardoso:2004nk,Kunduri:2006qa}. For the positive cosmological constant, the Kerr-dS black hole also has superradiance instability\cite{Zhang:2014kna}. Non-perturbative instabilities can be tested in various black holes. One of these instabilities is fragmentation instability, which was first given in Myers-Perry (MP) black holes\cite{Emparan:2003sy}. The fragmentation instability suggests that the rapidly rotating black hole can be broken into black holes due to centrifugal force. For MP black holes, the power of the gravitational force changes in higher dimensions. Thus, it can be thermodynamically possible, and we can predict instability in its rapid rotations. In addition, the MP black hole has instability in rapid rotations\cite{Shibata:2009ad,Dias:2009iu,Dias:2010eu,Dias:2010maa,Durkee:2010qu,Murata:2012ct,Dias:2014eua}, so fragmentation instability can act as a a guide for the existence of other kinds of instability. Fragmentation instability is obtained in Gauss-Bonnet and AdS black holes\cite{Gwak:2014xra,Ahn:2014fwa,Gwak:2015ysa}.

We will investigate the spin interaction with a cosmological constant in the collision of Kerr-(A)dS black holes. To relate the energy of the spin interaction to that of the gravitational radiation released under the collision, we suppose that one of the black holes is very small and slowly rotating. Then, the energy of the radiation will be thermodynamically obtained and proportional to the angular momentum both black holes. The source of the radiation will be found under the assumption that the small black hole is, approximately, a spinning particle coming into the other black hole. We can analytically obtain the potential of the spin interaction between the black hole and the spinning particle, and the potential will be identical to the energy of the radiation. Therefore, we can prove the energy of the spin interaction with the cosmological constant released in the collision of two black holes. We will expand our results to show the overall behaviors of the upper bounds on the radiation with respect to the angular momentum and mass of the black holes. In this analysis, the upper bounds on the radiation are affected by the instability of the final black hole after the collision.

This paper is organized as follows. We introduce the Kerr-(A)dS black hole and its thermodynamics in section~\ref{sec2}. In section~\ref{sec3}, we obtain the energy of the gravitational radiation generated from the collision of the Kerr-(A)dS black holes, and then the energy is proven to have the exact potential of the spin interaction for the system of the black hole and spinning particle. Then, the upper bounds of the radiation are given with respect to the mass and angular momentum of the black holes in section~\ref{sec4}. In section~\ref{sec5}, we briefly summarize our results.

\section{The Kerr-(A)dS Black Hole}\label{sec2}
The Kerr-(A)dS black hole is a solution to Einstein gravity with a cosmological constant in four-dimensional spacetime. The sign of the cosmological constant determines that of the curvature of the spacetime. We will analyze both positive and negative cases. The black hole is a rotating solution given in terms of mass and spin parameters $M$ and $a$,
\begin{align}\label{eq:metricKdS}
&ds^2=-\frac{\Delta_r}{\rho^2}\left(dt-\frac{a\sin^2\theta}{\Xi} d\phi\right)^2+\frac{\rho^2}{\Delta_r}dr^2+\frac{\rho^2}{\Delta_\theta}d\theta^2+\frac{\Delta_\theta\sin^2\theta}{\rho^2}\left(a\,dt-\frac{r^2+a^2}{\Xi}d\phi\right)^2\,,\\
&\rho^2=r^2+a^2\cos^2\theta\,,\,\,\Delta_r=(r^2+a^2)(1-\frac{1}{3}\Lambda r^2)-2mr\,,\,\,\Delta_\theta=1+\frac{1}{3}\Lambda a^2 \cos^2\theta\,,\,\,\Xi=1+\frac{1}{3}\Lambda a^2\,,\nonumber
\end{align}
where the coefficient $\Xi$ is defined as a positive value, so that it gives a BPS-like bound for the AdS cases\cite{Chrusciel:2006zs}. The mass and angular momentum $M_B$ and $J_B$ are\cite{Dolan:2013ft,Caldarelli:1999xj}
\begin{eqnarray}
M_B=\frac{m}{\Xi^2}\,,\quad J_B=\frac{ma}{\Xi^2}\,.
\end{eqnarray} 
The thermal properties of the black hole are defined on its horizons, and the definitions of these properties are consistent with all values of the cosmological constant. In this paper, we will mainly focus on the properties of the outer horizon $r_h$. The Hawking temperature $T_H$ and Bekenstein-Hawking entropy $S_{BH}$ are given as
\begin{eqnarray}
T_H=\frac{r_h\left(1-\frac{\Lambda a^2}{3}-\frac{a^2}{r_h^2}-\Lambda r_h^2\right)}{4\pi(r_h^2+a^2)}\,,\quad S_{BH}=\frac{\pi(r_h^2+a^2)}{\Xi}\,.
\end{eqnarray}
The angular velocity is calculated to $\Omega_h$ on the outer horizon, but the velocity includes the rotation of the coordinates. This can be shown by $M=0$ in Eq.~(\ref{eq:metricKdS}), where the rotation still exists\cite{Caldarelli:1999xj,Akcay:2010vt}. To remove this effect, we can determine the referential angular velocity $\Omega_\infty$ at $r\rightarrow \infty$. Hence, the angular velocity of the black hole $\Omega$ is\cite{Caldarelli:1999xj}
\begin{eqnarray}
\Omega=\Omega_h-\Omega_\infty=\frac{a\Xi}{r_h^2+a^2}-\frac{\Lambda a}{3}=\frac{a\left(1-\frac{\Lambda }{3}r_h^2\right)}{r_h^2+a^2}\,,
\end{eqnarray}
which is consistent with an arbitrary cosmological constant. On the outer horizon, the first law of thermodynamics is defined as
\begin{eqnarray}
dM_B=\Omega dJ_{B} + T_H dS_{BH}\,.
\end{eqnarray}
In the case of a dS spacetime, the spacetime has the cosmological horizon $r_c$ outside of the outer horizon. Compared with the AdS or flat case, the observable region is confined in a dS case. The first law of thermodynamics on the cosmological horizon is defined in the same way as that of the outer horizon, so that
\begin{eqnarray}
dM_B=\Omega_c dJ_{B} + T_c dS_{c}\,,
\end{eqnarray}
where the temperature $T_c$, entropy $S_c$, and angular velocity $\Omega_c$ on the cosmological horizon are
\begin{eqnarray}
T_c=\frac{r_h\left|1-\frac{\Lambda a^2}{3}-\frac{a^2}{r_c^2}-\Lambda r_c^2\right|}{4\pi(r_c^2+a^2)}\,,\quad S_{c}=\frac{\pi(r_c^2+a^2)}{\Xi}\,,\quad \Omega_c=\frac{a\left(1-\frac{\Lambda }{3}r_c^2\right)}{r_c^2+a^2}\,.
\end{eqnarray}
The absolute of the temperature is for the positive definite because the surface gravity of the cosmological horizon is negative.

\section{Spin Interaction and Gravitational Radiation}\label{sec3}
We will show that the energy of the spin interaction between two Kerr-(A)dS black holes contributes to the gravitational radiation released upon their collision. We will, thermodynamically, yield the energy of the radiation when one of the black holes has a tiny mass and small angular momentum to be considered a spinning particle so that higher-multipole moments can be ignored. Then, the energy of the radiation will exactly correspond to the potential of the spin interaction from the Mathisson-Papapetrou-Dixon (MPD) equation for the spinning particle near the black hole. In this analysis, we assume that the effects from the cosmological constant, such as the AdS boundary and cosmological horizon, are governed by the more massive of the two black holes.

\subsection{The Gravitational Radiation in the Collision of Two Kerr-(A)dS Black Holes}\label{sec31}
The energy resulting from gravitational radiation is investigated under the collision of two Kerr-(A)dS black holes. We assume that one of the black holes is small because the radiation energy will meet the potential of the spin interaction in the system of the spinning particle and the black hole. To determine the energy of this radiation, we suppose that the first and second Kerr-(A)dS black holes with masses $M_1$ and $M_2$ stay far away from each other, and we assume that they are separated in the direction of their rotating axes so that their rotating planes are parallel\cite{Wald:1972sz}. Then, slowly moving from the gravitational attraction, the two black holes undergo a head-on collision and merge into the final black hole, which has mass $M_f$. The difference between the masses of the initial and final states can express the gravitational radiation, $M_{r}$, under 
\begin{eqnarray}
M_r=(M_1+M_2)-M_f\,,
\end{eqnarray}
where we only consider the radiation $M_r$ that was swept the first time because the radiation energy will be matched to the energy of the spin interaction, so that we do not need to take account of a reflection at the AdS boundary. To obtain the energy of radiation $M_r$, we assume that the angular momentum is conserved as
\begin{eqnarray}
J_1+J_2 = J_f\,,
\end{eqnarray}
where the angular momenta of the first and second black holes are $J_1$ and $J_2$, and the final one is $J_f$. Since the head-on collision of the initial black holes occurs from gravitational attraction, the Bekenstein-Hawking entropy of the final black hole $S_f$ should be greater than the sum of those of the first and second black holes, $S_1$ and $S_2$. Hence,
\begin{eqnarray}\label{eq:ent09}
S_1+S_2\leq S_f\,.
\end{eqnarray} 
The radiation energy depends on the initial condition of the black holes. To be precise, using the given initial conditions of the black holes, which are described as $m_1$ and $a_1$ for the first one and $m_2$ and $a_2$ for the second one, we can obtain the upper bound on the radiation and the lower bound on the mass parameter $m_f$ of the final black hole with the spin parameter $a_f$. To describe the radiation in terms of the second black holes parameters, we fix the first black hole and suppose the second black hole to have a small mass and angular momentum as compared with the first one, so $M_1\gg M_2$ and  $M_1^2\gg J_2$. We can calculate the maximum value of the radiation using the equality of Eq.~(\ref{eq:ent09}). The change of the maximum radiation $M_{max}$ can be given in terms of the partial derivative with respect to the angular momentum of the second black hole, so the leading term is
\begin{eqnarray}\label{eq:partial01}
\frac{\partial M_{max}}{\partial J_2}=-\frac{2m_1a_1r_1}{(r_1^2+a_1^2)^2}+\mathcal{O}(J_2)\,.
\end{eqnarray}
From Eq.~(\ref{eq:partial01}), the potential of the spin interaction $U_{s}$ can be obtained and rewritten in terms of the cosmological constant
\begin{eqnarray}\label{eq:spin02}
U_{s}=\frac{\left(1-\frac{1}{3}\Lambda r_1^2\right)^2\Xi_1^2}{2m_1^2 r_1}J_1 J_2\,,
\end{eqnarray}
which is also the maximum value of the radiation. For both black holes, the overall sign of the potential only depends on the rotating direction. Under a parallel arrangement, the potential is positive, so the black holes are repulsive to each other. In addition, an anti-parallel arrangement gives a negative potential so that the two black holes will be attracted by the  spin interaction. The potential is now an effective value, but it will correspond exactly to the potential obtained from the MPD equation for a spinning particle near the black hole. This correspondence will be explored in a following section.

\subsection{The Potential of Spin Interaction from the MDP Equation}\label{sec32}

Since the second black hole is small compared with the first one, we can treat the second black hole as though it is in the limit of a spinning particle. We show that the potential of the spin interaction in Eq~(\ref{eq:spin02}) corresponds exactly to that of the spinning particle near a Kerr-(A)dS black hole. The motions of the spinning particle are governed under the MPD equation\cite{Mathisson:1937zz,Papapetrou:1951pa,Dixon:1970zza} with linear momentum $p^\mu$ and velocity $u^\mu$ so that
\begin{eqnarray}
\frac{Dp^\mu}{Ds}=-\frac{1}{2}R^\mu_{\nu\rho\sigma}u^\nu S^{\rho\sigma}\,, \quad \frac{DS^{\mu\nu}}{Ds}= p^\mu u^\nu - p^\nu u^\mu\,,
\end{eqnarray}
where the spin tensor and the proper time are $S^{\mu\nu}$ and $s$, respectively, in the spacetime of the Riemann curvature tensor $R^\mu_{\nu\rho\sigma}$. The trajectory of the spinning particle can be properly described under the choice of a supplementary condition\cite{Beiglbock1967}
\begin{eqnarray}
p_\mu S^{\mu\nu} = 0\,.
\end{eqnarray}
The spinning particle has the mass $\mu$ and the magnitude of the spin $S$ in terms of the spin tensor and linear momentum
\begin{eqnarray}
S^2 = \frac{1}{2}S_{\mu\nu}S^{\mu\nu}\,, \quad  \mu^2 =-p_\mu p^{\mu}\,.
\end{eqnarray} 
Now, the spinning particle is presumed to be the second black hole in section~\ref{sec31}, so the spin monopole of the particle is only taken account of in the MDP equations. Hence, the momentum of the particle is proportional to the velocity
\begin{eqnarray}
\frac{DS^{\mu\nu}}{Ds} = 0\,, \quad p^\mu = \mu v^\mu\,.
\end{eqnarray}
Since the direction of the head-on collision is pole-to-pole, the spinning particle also comes into the north pole of the first black hole in the initial condition, and its spinning plane is also parallel to that of the black hole. Then, the normalized velocity and spin vectors are set up
\begin{eqnarray}
v^\mu = \left(\dfrac{1}{\sqrt{-g_{tt}}},v^r,0,0 \right), \qquad \text{and} \qquad S^{\mu} = \left(0, \dfrac{1}{\sqrt{g_{rr}}} S, 0,0\right)\,,
\end{eqnarray}
where the components of the metric are that of the Kerr-(A)dS black hole. The kinetic energy of the spinning particle is not a conserved quantity as the energy now includes the potential of the spin interaction. And, since the potential of the spin interaction depends on the spin of the particle, the potential can be changed along with the motion of the particle. The conserved quantities $C_\xi$ for Killing vector $\xi^\mu$ are given\cite{Dixon:1970zza} as
\begin{eqnarray}
C_\xi = p_\mu \xi^\mu + \dfrac{1}{2} S^{\mu\nu} \nabla_\mu \xi_\nu\,,
\end{eqnarray}
where the energy of the particle is for the Killing vector $\xi^\mu_{(t)} = \delta^\mu_{t}$, then
\begin{eqnarray}\label{eq:energy03}
E = -p_t - \dfrac{1}{2} S^{\mu\nu} \nabla_\mu g_{\nu t}\,,
\end{eqnarray}
where the potential of the spin interaction can be directly obtained. In the energy of Eq.~(\ref{eq:energy03}), the first term refers to the kinetic energy of the linear momentum, and the second term represents the potential of the spin interaction. The radiation will be contributed to the potential of the spin interaction at the outer horizon, so the potential should be yielded at the outer horizon of the first black hole
\begin{eqnarray}
E_s = \frac{2m_1a_1r_1}{(r_1^2+a_1^2)^2}S+\frac{a\Lambda}{3}S\,,
\end{eqnarray}
in which the second term is from the rotation of the spacetime. This term can be removed in consideration of referential angular velocity $\Omega_\infty$\cite{Caldarelli:1999xj}. In the energy of the particle, the effect of this velocity appears as additional energy, which is proportional to the angular momentum\cite{Gwak:2015fsa}. The normalized potential of the spin interaction is obtained as
\begin{eqnarray}
U_s = \frac{\left(1-\frac{1}{3}\Lambda r_1^2\right)^2\Xi_1^2}{2m_1^2 r_1}J_1 S\,,
\end{eqnarray}
which is exactly that of Eq.~(\ref{eq:spin02}) under $S=J_2$. Therefore, in this limit for the black holes, the potential of the spin interaction is released as gravitational radiation. In addition, the parallel arrangement interacts with the repulsion of the black holes, and the anti-parallel arrangement interacts with the attraction. Therefore, we can expect that the energy of the radiation is greater for the anti-parallel arrangement than for the parallel one. To confirm this behavior, we apply the thermodynamic approach of section~\ref{sec31} without the limit of the second black hole.

\section{Upper Bound on the Radiation under the Collision}\label{sec4}
Here, we expand our results from section~\ref{sec31} to illustrate approximately the trend of radiation with respect to mass and angular momentum using thermodynamic analysis. In the case of large mass and angular momentum, the effects of the cosmological constant become greater than before, so we cannot ignore these under the collisions. In this analysis, we simply show the upper bound on the radiation with respect to the mass and angular momentum under dimensionless coordinates scaled by $\sqrt{\lambda/\Lambda}$, in which $\lambda$ is $-1$ for AdS and $+1$ for dS cases. Then,
\begin{align}\label{eq:dimless}
&\tilde{s}=\frac{s}{\sqrt{\lambda/\Lambda}}\,,\quad\tilde{t}=\frac{t}{\sqrt{\lambda/\Lambda}}\,,\quad\tilde{r}=\frac{r}{\sqrt{\lambda/\Lambda}}\,,\quad\tilde{M}=\frac{M}{\sqrt{\lambda/\Lambda}}\,,\quad\tilde{a}=\frac{a}{\sqrt{\lambda/\Lambda}}\,,\\
&\tilde{\rho}^2=\tilde{r}^2+\tilde{a}^2\cos^2\theta\,,\,\,\tilde{\Delta}_{\tilde{r}}=(\tilde{r}^2+\tilde{a}^2)(1-\frac{1}{3}\lambda \tilde{r}^2)-2\tilde{m}\tilde{r}\,,\,\,\tilde{\Delta}_\theta=1+\frac{1}{3}\lambda \tilde{a}^2 \cos^2\theta\,,\,\,\tilde{\Xi}=1+\frac{1}{3}\lambda \tilde{a}^2\,,\nonumber
\end{align}
and tildes are omitted for simplicity. The upper bounds on the radiation are given approximately through the thermodynamic approach without considering the effect of the cosmological constant, so the upper bounds need to study about their possibilities. However, the trends of the radiation can be approximately shown in this approach. In addition, interesting behaviors related between the gravitational radiation and stability of the black hole can be discussed under this method.

The general behaviors for an arbitrary cosmological constant are similar to each other, as shown in Fig.~\ref{fig:f1}. As we have already obtained in Eq.~(\ref{eq:spin02}), the derivative of the bounds on the radiation with respect to the angular momentum of the second black hole is negative. So, the bounds on the radiation are greater in anti-parallel arrangements of black holes than in parallel arrangements for any values of the cosmological constants in Fig~\ref{fig:f1}~(a).
{\begin{figure}[h]
\centering\subfigure[{The bounds on the radiation with respect to the angular momentum of the second black hole. The black holes are $m_1=0.1$, $a_1=0.05$, and $m_2=0.1$\,.}]
{\includegraphics[scale=0.70,keepaspectratio]{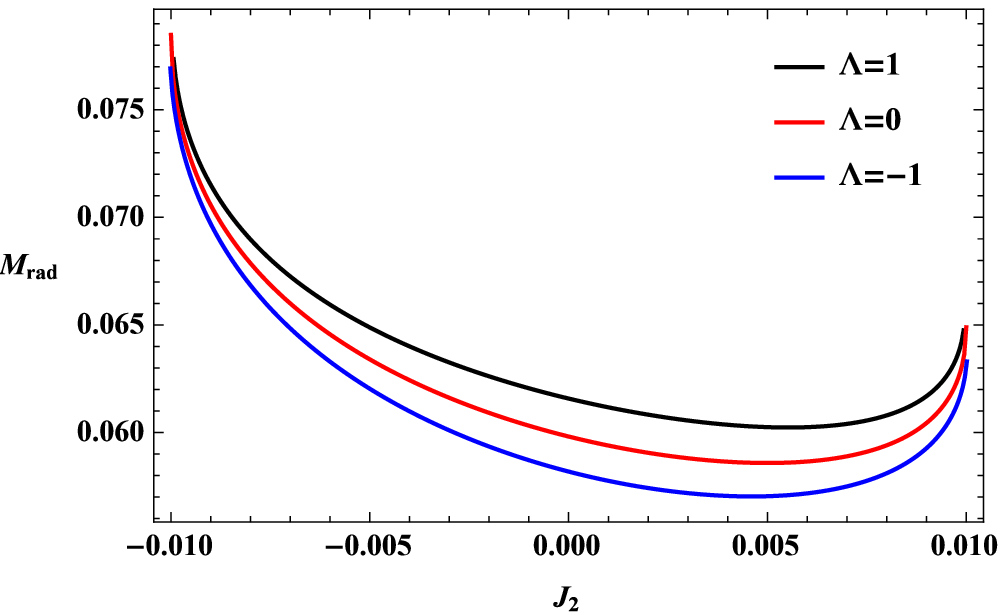}}
\quad
\centering\subfigure[{The bounds on the radiation with respect to the mass of the second black hole. The black holes are $m_1=0.1$, $a_1=0.05$, and $a_2=0.05$\,.}]
{\includegraphics[scale=0.70,keepaspectratio]{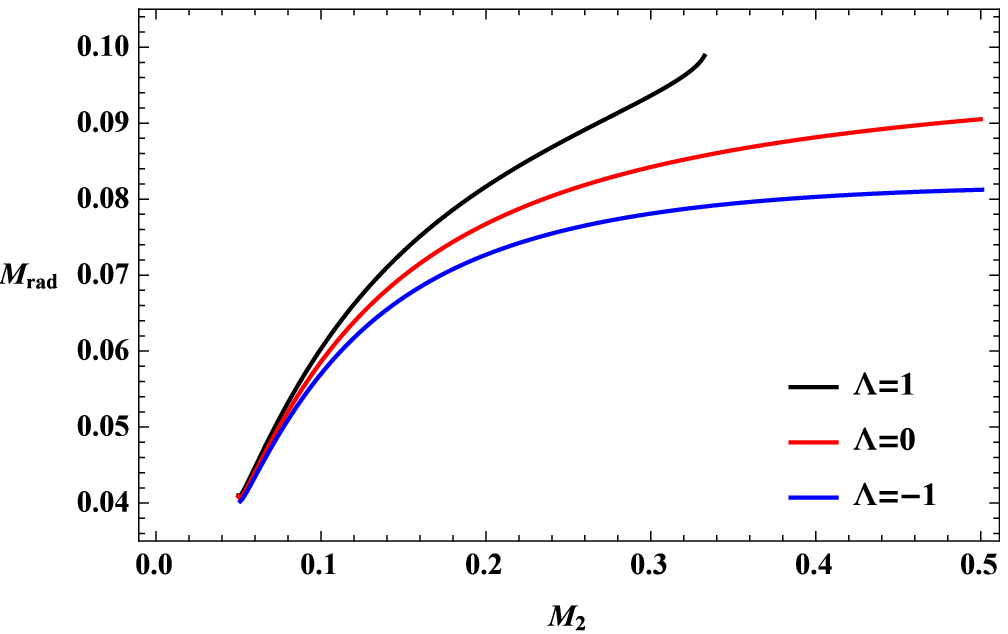}}
\caption{{\small The bounds on the radiation for the collisions of two KdS black holes.}}
\label{fig:f1}
\end{figure}}
The radiation in the anti-parallel arrangement includes the contribution from the negative potential of the spin interaction between the black holes. For this reason, the minimum radiation energy appears in the parallel arrangement having the positive potential. Radiations are greater for the positive cosmological constant than for the negative one, as shown in Fig.~\ref{fig:f1}~(a). The bounds on the radiation increase along with the mass of the second black hole, as shown in Fig.~\ref{fig:f1}~(b). For the fixed angular momentum, the minimum masses of the black holes are given from the extremal condition. Consequently, the bounds start at the finite value of the mass in Fig.~\ref{fig:f1}~(b). Then, the bounds on radiation are proportional to the mass of the second black hole, but the dS and AdS cases have an upper limit of radiation due to the effects of the cosmological constant.

For a positive cosmological constant, the overall behaviors are similar to the cases for Kerr black holes. The parameter range in which the Kerr-dS black hole exists is very limited due to the extremal condition and the cosmological horizon, resulting in a narrow range for the angular momentum of the radiation, as shown in Fig.~\ref{fig:f2}~(a).
{\begin{figure}[h]
\centering\subfigure[{The bounds on the radiation with respect to the angular momentum of the second black hole. The black holes are $m_1=0.1$ and $a_1=0.05$ with the positive cosmological constant.}]
{\includegraphics[scale=0.70,keepaspectratio]{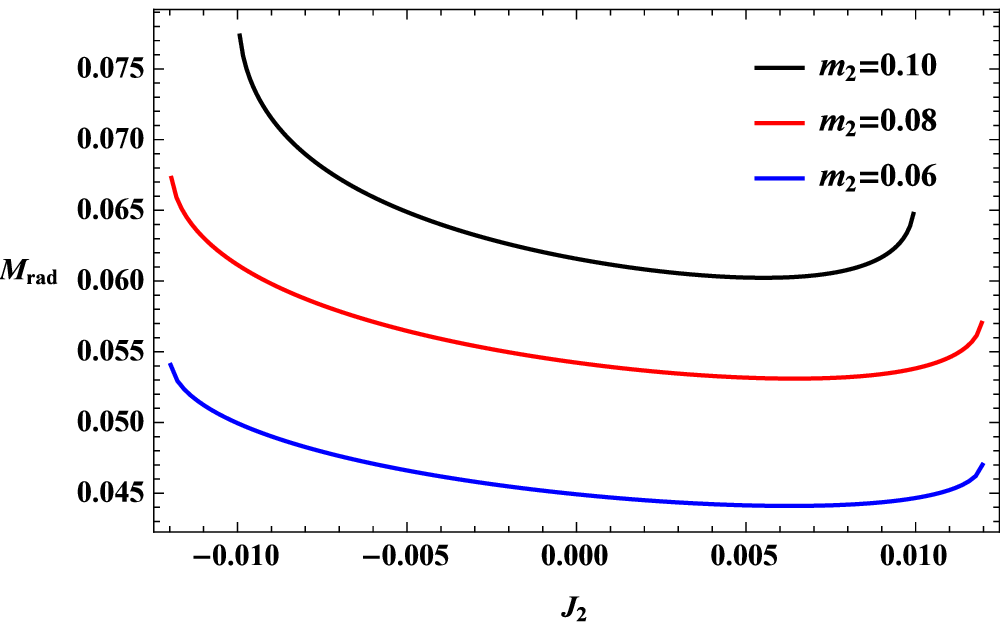}}
\quad
\centering\subfigure[{The bounds on the radiation with respect to the mass of the second black hole. The black holes are $m_1=0.1$ and $a_1=0.05$ with the positive cosmological constant.}]
{\includegraphics[scale=0.70,keepaspectratio]{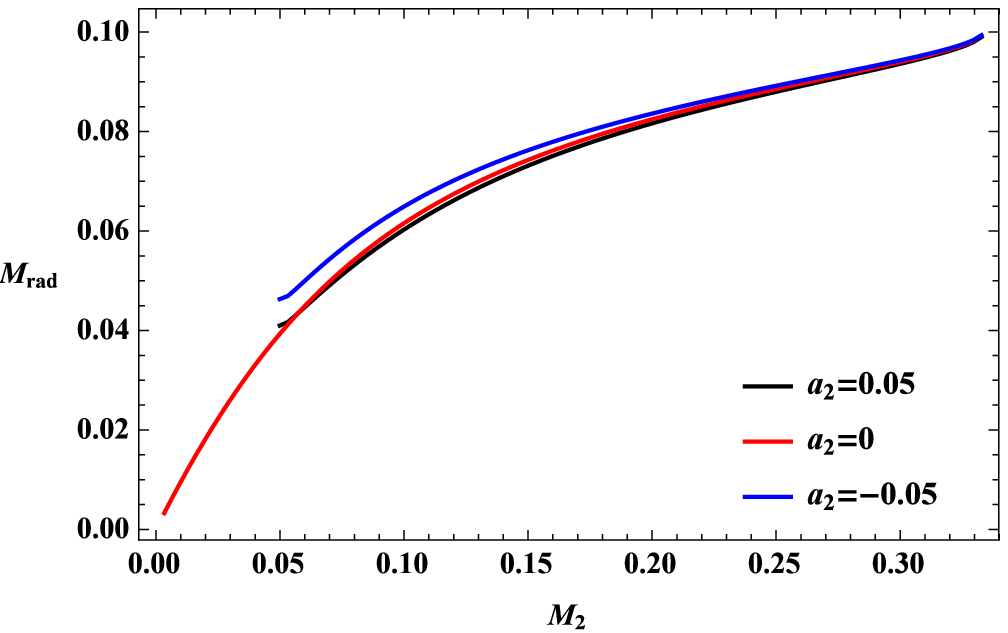}}
\caption{{\small The bounds on the radiation for the collisions of two KdS black holes.}}
\label{fig:f2}
\end{figure}}
In Fig.~\ref{fig:f2}~(a) and (b), the bounds on the radiation normally increase in accordance with the mass of the second black hole, and more radiation can be released in near-extremal black holes. The Kerr-dS black hole with a spin parameter of zero can have an outer horizon with a small mass, so radiation can be observed in the cases of the second small black hole. In addition, there are minimum masses of the second black hole that can release the radiation for given spin parameters due to the extremal condition in Fig.~\ref{fig:f2}~(b).

The bound on the collision of the Kerr-AdS black holes is different from that of flat or dS ones due to the effect of the negative cosmological constant. With respect to the angular momentum of the second black hole, the radiation is greater in an anti-parallel arrangement than in a parallel one, as shown Fig.~\ref{fig:f4}~(a).
{\begin{figure}[h]
\centering\subfigure[{The bounds on the radiation with respect to the angular momentum of the second black hole. The black holes are $m1=0.1$ and $a1=0.05$ with the negative cosmological constant.}]
{\includegraphics[scale=0.70,keepaspectratio]{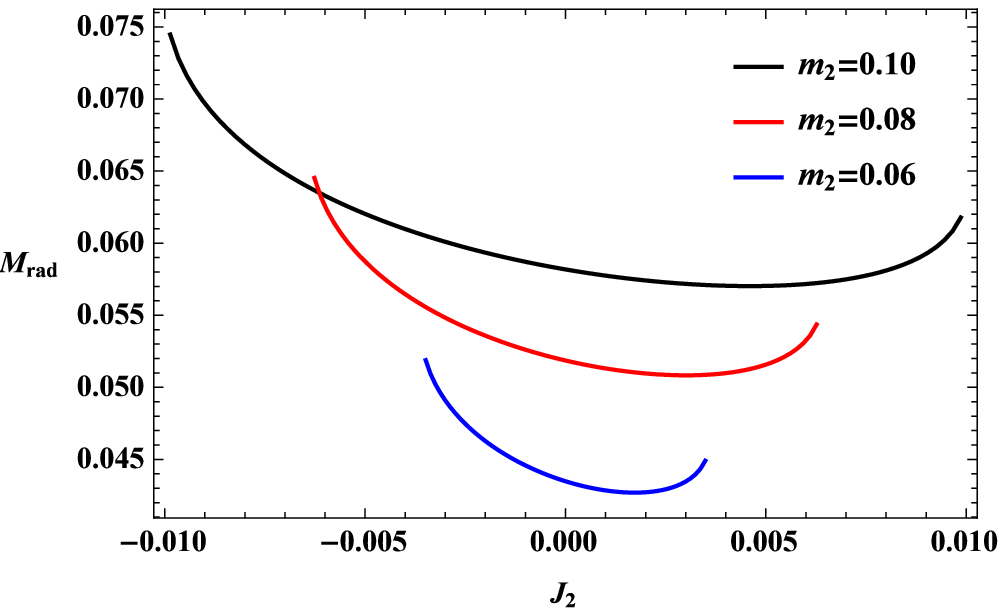}}
\quad
\centering\subfigure[{The bounds on the radiation with respect to the mass of the second black hole. The black holes are $m1=0.1$ and $a1=0.05$ with the negative cosmological constant.}]
{\includegraphics[scale=0.70,keepaspectratio]{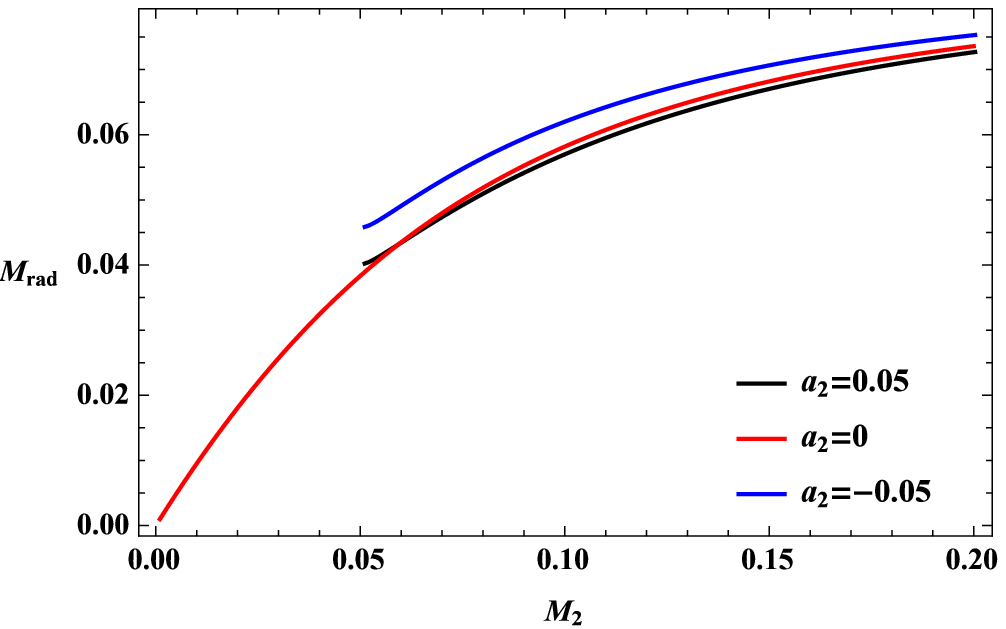}}
\caption{{\small The bounds on the radiation for the collisions of two KAdS black holes.}}
\label{fig:f4}
\end{figure}}
For the second small black hole, the radiation increases along with the mass of the second black hole, as shown in Fig.~\ref{fig:f4}~(b).
{\begin{figure}[h]
\centering\subfigure[{The bounds on the radiation with respect to the angular momentum of the second black hole. The black holes are $m1=1.0$ and $a1=0.25$ with the negative cosmological constant.}]
{\includegraphics[scale=0.70,keepaspectratio]{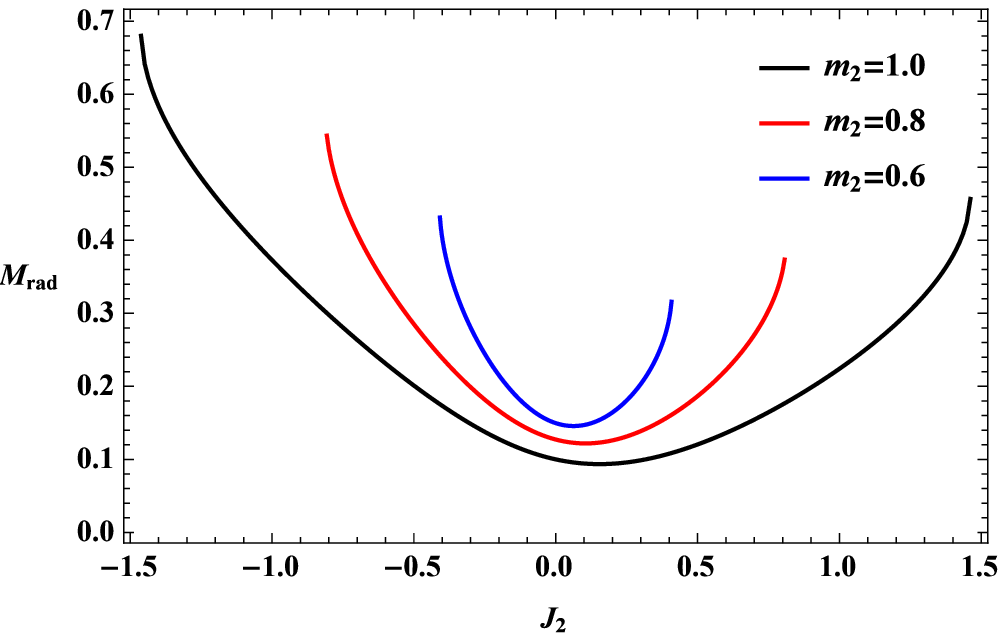}}
\quad
\centering\subfigure[{The bounds on the radiation with respect to the mass of the second black hole. The black holes are $m1=1.0$ and $a1=0.25$ with the negative cosmological constant.}]
{\includegraphics[scale=0.70,keepaspectratio]{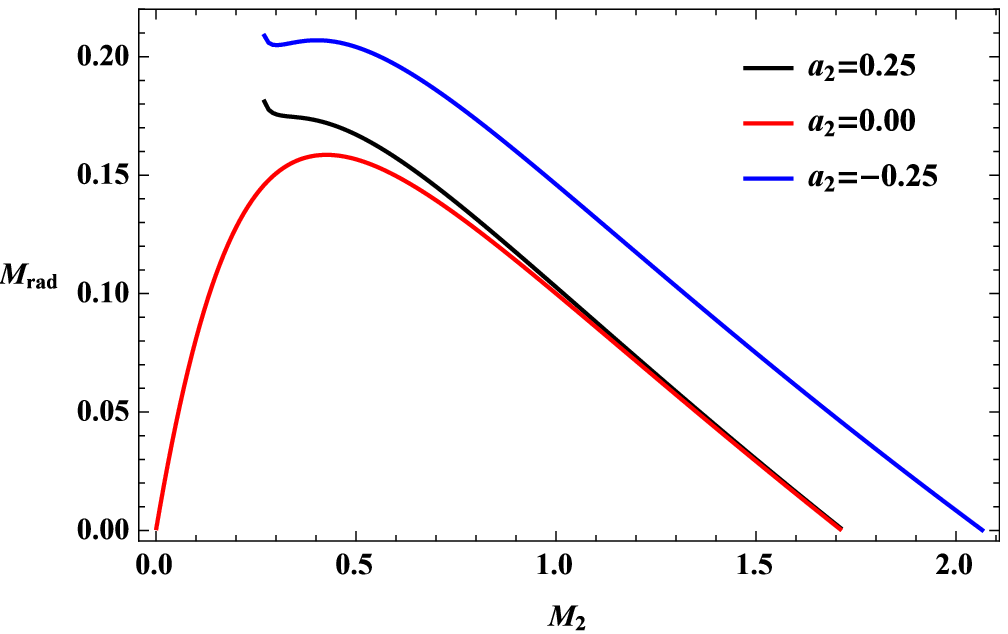}}
\caption{{\small The bounds on the radiation for the collisions of two KAdS black holes.}}
\label{fig:f3}
\end{figure}}
However, for the second black hole that has a large mass, the radiation can decrease along to the increase of the mass of the second black hole, as shown in Fig.~\ref{fig:f3}~(a). However, for the large mass of the second black hole, the radiation is diminished until it finally becomes zero, as shown Fig.~\ref{fig:f3}~(b). Zero radiation, approximately, can be expected in an analytical method. To demonstrate this behavior, we suppose the collision of two large Schwarzschild-AdS black holes with the same mass. In this way, the outer horizon of the initial black hole is $r_1=r_2\sim (6m)^{1/3}$, and the outer horizon of the final black hole is $r_f\sim \sqrt{2} (6m)^{1/3}$. The difference between the masses are within the limit of $m\gg 1$
\begin{eqnarray}
M_{rad}=(M_1+M_2) -M_f=(2-2\sqrt{2})m<0\,.
\end{eqnarray}
Therefore, no radiation is released in the collision of masses that are large enough. This result was attained without consideration of the AdS effect on the initial condition. Although more detailed analysis may be required, a radiation of zero may originate from the effect of instability in the Kerr-AdS black hole. The instability of Kerr-AdS black holes can increase according to the angular momentum. The fragmentation instability based on thermodynamics especially shows similar results to a reverse process of the collision\cite{Gwak:2014xra}. This implies that there is no more energy that can be released as radiation because too much energy is needed to synthesize the unstable final state. Therefore, the instability of the black hole can affect the radiation released in the collision of Kerr-AdS black holes.

\section{Summary}\label{sec5}
We have investigated spin interaction and its relation to radiation under the collision of the Kerr-(A)dS black holes. Using a thermodynamic approach, we have obtained the energy of the gravitational radiation under the collision of black holes by assuming that one has a very small mass and angular momentum. Then, the energy of the radiation has been found to be exactly that of the potential of the spin interaction between the black hole and a spinning particle in the limit. Therefore, the energy of the spin interaction can be released under the collision. Depending on the cosmological constant, the spin interaction appears as an attraction for an anti-parallel arrangement and as a repulsion for a parallel one. To discover the overall phenomena regarding the radiation, we simply expanded thermodynamic approach to arbitrary mass and angular momentum and obtained the upper bounds on the radiation instead of the exact expressions. In the most cases, radiation may be greater in the anti-parallel arrangement than in the parallel one, and it may be proportional to the masses of the black holes in the initial state due to the spin interaction. The cases of the Kerr-dS are similar to those of the Kerr black holes, and more energy from radiation is expected for a given mass when it is compared with flat cases. However, radiation under the collision of Kerr-AdS black holes can be closely related to the instability of the black hole in the final state. The increase of instability can diminish the energy of radiation under the collision because more energy is needed to synthesize an unstable final state than a stable one.

\vspace{10pt} 

{\bf Acknowledgments}

BG was supported by Basic Science Research Program through the National Research Foundation of Korea(NRF) funded by the Ministry of Science, ICT \& Future Planning(NRF-2015R1C1A1A02037523). DR was supported by the Korea Ministry of Education, Science and Technology, Gyeongsangbuk-Do and Pohang City.

\bibliographystyle{mod}
\bibliography{bib}
\end{document}